\documentclass{article}

\usepackage{arxiv}

\usepackage[utf8]{inputenc}            
\usepackage[T1]{fontenc}               
\usepackage[hidelinks]{hyperref}       
\usepackage{url}                       
\usepackage{booktabs}                  
\usepackage{amsfonts}                  
\usepackage{nicefrac}                  
\usepackage{microtype}                 
\usepackage{lipsum}		               
\usepackage{graphicx}
\usepackage[numbers,sort&compress]{natbib}
\usepackage{doi}
\usepackage{caption}

\title{Material Experience: An Evaluation Model for Creative Materials Based on Visual-Tactile Sensory Properties}

    \author{ 
        \includegraphics[scale=0.06]{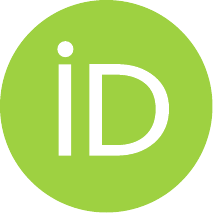}\hspace{1mm}Yuxin Zhang
        \thanks{Yuxin Zhang, PhD in Design, focuses on industrial design, material experience, affective engineering, and ergonomics. For further academic details, please refer to his ORCID profile: https://orcid.org/0000-0001-8100-3649} \\
        Academy of Arts \& Design \\
        Tsinghua University \\
        Haidian  District, Beijing, 100084, China \\
        \texttt{zhangyuxin@mail.tsinghua.edu.cn} \\
    \And
	Fan Zhang \\
	Academy of Arts \& Design \\
	Tsinghua University \\
	Haidian  District, Beijing, 100084, China \\
	\texttt{zhang-f24@mails.tsinghua.edu.cn} \\
    \And
	Jinjun Xia \\
        School of Arts \\
        Chongqing University \\
        Shapingba District, Chongqing, 400044, China \\
        \texttt{design@cqu.edu.cn} \\
    \And
	Chao Zhao \\
        Academy of Arts \& Design\\
        Tsinghua University\\
        Haidian District, Beijing, 100084, China \\
        \texttt{zhaochao@tsinghua.edu.cn} \\
}

\renewcommand{\shorttitle}{EVALUATION MODEL FOR CREATIVE MATERIALS}

\hypersetup{
pdftitle={A template for the arxiv style},
pdfsubject={q-bio.NC, q-bio.QM},
pdfauthor={David S.~Hippocampus, Elias D.~Striatum},
pdfkeywords={First keyword, Second keyword, More},
}

\begin{document}
\maketitle

\begin{abstract}
	This study adopts a design-oriented approach to integrate traditional braids with commonly used matrix materials, developing creative materials with different sensory properties by altering matrix material types and braid patterns. Based on these creative materials, a quantitative and structured model is proposed to assist designers understanding the material experience process and guide material selection by analyzing the relationship between material properties and sensory perception. Specifically, participants evaluated the creative materials under visual-tactile conditions using a 7-point semantic differential (SD) scale. Correlation analysis was performed to explore the data. The main and interaction effects of matrix materials and braid patterns on impression evaluation were analyzed using two-way analysis of variance (ANOVA). A structural equation model (SEM) was constructed based on exploratory factor analysis (EFA), and path coefficients were computed to assess the relative importance of material properties in determining material attractiveness. The results show that, compared to braids, the creative materials resulted in significant changes in impression evaluation. Furthermore, the creative materials can be understood through intrinsic, aesthetic, and physical properties, with their standardized regression coefficients for material attractiveness of 0.486, 0.650, and 0.103, respectively. These properties are interrelated and under their combined influence affect the attractiveness of the material. Therefore, designers should consider utilizing these relationships to enhance sensory experience in order to achieve design objectives. Moreover, designers should also consider balancing technology and experience, using materials according to the principle of "form follows function".
\end{abstract}


\section{Introduction}
    Experience design has emerged as one of the most prominent research topics in recent years, with satisfying experiences not only driving repeat purchase behavior \cite{oliverCognitiveAffectiveAttribute1993}, but also eliciting positive emotional responses \cite{hassenzahlDesigningMomentsMeaning2013}. In product design, the experience that materials evoke during user interactions are essential to understanding the design process. Consequently, materials are often seen as the starting point of the creative process and have the potential to be exploited to evoke a unique and meaningful product experience \cite{karanaMaterialsExperience2015a, pedgleyMaterialsExperienceFoundation2016}. However, a great deal of thought in the Human-Computer Interaction (HCI) community has been focused on addressing functionality and usability, with relatively little attention has been paid to the clear quality of experience in design \cite{lowgrenFluencyExperientialQuality2007}. Until recently, a growing number of HCI scholars have begun to use materials as an entry point to understand the possibilities of materials from an experience design perspective \cite{giaccardiFoundationsMaterialsExperience2015}.
    
    Despite this, there are still numerous challenges in the selection and utilization of materials \cite{ashbySelectionStrategiesMaterials2004}. To date, little is known about the processes through which humans perceive materials, and a systematic approach to defining and designing material experiences has yet to be established \cite{karanaMaterialDrivenDesign2015}. Moreover, material perception is highly multimodal or multisensory, involving a combination of vision, hearing, touch, smell, and taste to varying degrees \cite{guestAudiotactileInteractionsRoughness2002, martinMultimodalPerceptionMaterial2015}. As a result, applying knowledge of materials to quantitative analysis and aesthetic computing is a complex task. The systematic study of material perception not only enables researchers and designers to understand how to experience a particular material and improve it accordingly for commercial success, but also inspires designers and material developers to come up with create innovative material and product concepts \cite{karanaMaterialsExperienceFundamentals2014}. These trends also provide an opening for the emergence of computational aesthetic.
    
    Humans seem to be rather good at making judgments about material classes \cite{flemingPerceptualQualitiesMaterial2013} and material properties \cite{wiebelSpeedAccuracyMaterial2013}. Sufficient research suggests that a "material turn" \cite{roblesTexturingMaterialTurn2010} is taking place within interaction design, leading to the establishment of a "material strategy" \cite{vallgardaMaterialStrategyExploring2010}. In the 1960s, Tsutomu Suzuki proposed the concept of "Materials Planning" which advocated for understanding the characteristics of materials from both physical and sensory perspectives and emphasized the role of human factors in design \cite{suzukiFrameworkMaterialsPlanning2005}. Therefore, designers must recognize the duality of product materials, which satisfy both functional and hedonic needs \cite{hassenzahlExperienceDesignTechnology2010}. It is also evident that the development and application of new materials must be multidisciplinary \cite{hornbuckleMobilizingMaterialsKnowledge2021}. On the other hand, Karana et al. \cite{karanaMaterialsExperienceFundamentals2014} introduced the term "material experience" and emphasized the important role materials play in shaping how people interact with and experience products. They conducted interviews and questionnaires with 20 industrial product designers, finding that designers primarily focus on sensory characterization during the material selection process. They also highlighted the importance of intangible characteristics of materials (ICM) such as perceived value, associated emotions and meaning. Therefore, product designers are responsible for selecting appropriate materials for their products by considering both the technical and sensorial characteristics of materials \cite{karanaMaterialConsiderationsProduct2008}. In summary, the exploration of material perception provides a comprehensive understanding of materials that can guide design practice.
    
    To some extent, designers can influence people's emotions when interacting with a product \cite{demirbilekProductDesignSemantics2003}. Based on the cognition of the material, the clever utilization of material experiences often leads to unexpected results. Relevant studies have shown that inconsistencies between prior expectations and posterior reality may affect perceptual quality of the material \cite{luddenVisualTactualIncongruities2009, luddenSurpriseLongitudinalStudy2012}. For instance, visual-tactile inconsistencies can lead to consumer surprise, satisfaction, delight, or disappointment. Yanagisawa \cite{yanagisawaEffectsVisualExpectation2015} demonstrated that different visual stimuli altered participants' responses to tactile samples. Liang \cite{liangDesigningUnexpectedEncounters2012} extended this concept to the field of digital materials and explored the interactive experience of human-centered digital materials. Therefore, understanding materials helps designers grasp the relationship between the physical properties of a surface as a design parameter and the customer's psychological response to the surface \cite{yanagisawaEffectsVisualExpectation2015}. Despite increasing number of research emphasizing the value of experiential concerns, the technology-experience imbalance continues to dominate material selection \cite{pedgleySpecialFileFutures2010}. Consequently, it is crucial for designers to consider the material selection from both a physical property and experiential perception.
    
    In the industrial sector, polymeric materials with excellent physical properties are widely used \cite{jiangExtrusion3DPrinting2020}, such as plastics and silicone. However, these materials often convey a perception of being inexpensive \cite{ljungbergMaterialsSelectionDesign2003}. To address this issue, the strategy of adding biomimetic fillers (e.g., wood or marble) into the matrix material has been employed \cite{appelsFabricationFactorsInfluencing2019}, although this approach has certain limitations. In contrast, fabrics have garnered attention as reinforcing materials when combined with matrix materials \cite{azwaReviewDegradabilityPolymeric2013}. Particularly, craft braids, as a form of fabric, not only provide aesthetically pleasing due to their regular and dynamic patterns, but also possess unique mechanical properties resulting from their distinctive woven structures. These properties can imbue composites materials with more meaningful characteristics \cite{ayranci2DBraidedComposites2008}. More importantly, they hold significant humanistic value, being integral to both society and culture. While studies on braids have been conducted, relatively few researches have focused on the changes in material experience following the compounding process. As the use of composites increases and research on mechanical properties expands, exploring the changes in sensory experiences when braids are combined with polymeric materials becomes increasingly important. Braid composites offer a more complex appearance and richer tactile experience compared to single materials, providing designers with a deeper understanding of industrial materials. Therefore, this study investigates the production method that combines traditional craft braids with polymer matrix materials to fabricate creative materials, with focus on the sensory experiences. Additionally, explores the role of creative materials in advancing the dissemination of cultural heritage.
    
    This study is based on the premise that general principles of aesthetic pleasure unified in human nature. That is, regardless of their varied manifestations, the aesthetic experience adheres to principles that apply universally \cite{berghmanUnifiedModelAesthetic2017}. In contrast, the representations of materials are diverse. Therefore, the study explores the optimal balance between unity and variety \cite{postPreserveUnityAlmost2016}, and incorporates this balance into a theoretical model. Furthermore, we focus on visual and tactile senses as the primary sensory modalities in material experience. The research subjects include braids and braid composites with different matrix materials. The purpose of this study is to clarify the relationship between material properties and sensory perception (Attractiveness of the Material), as well as to extend material experience theory, particularly with regard to the influence of different matrix materials and reinforcement materials on impression evaluation. Additionally, a generalized structural equation model is proposed to assist designers in the selection of materials.

\section{Experiment}
\subsection{Stimulus}
    The material of a craft is often a way to attract people's initial attention \cite{karanaMaterialsExperience2015a}. In this study, we used a traditional craft braid made by crossing eight yarns at a specific angle, called "Eight Vestiges", which holds both practical and artistic value \cite{zhangIMPRESSIONEVALUATIONBASED2024}. The braids originated in ancient China and have existed for thousands of years, representing a crystallization of human wisdom \cite{branscombNewDirectionsBraiding2013, careyIntroductionBraidedComposites2017}. During the braiding process, a braid disk was used to produce braid that helps beginners easily learn the manufacturing method and standardize production.

\begin{figure}[h]
\centering
\includegraphics[width=0.9\textwidth]{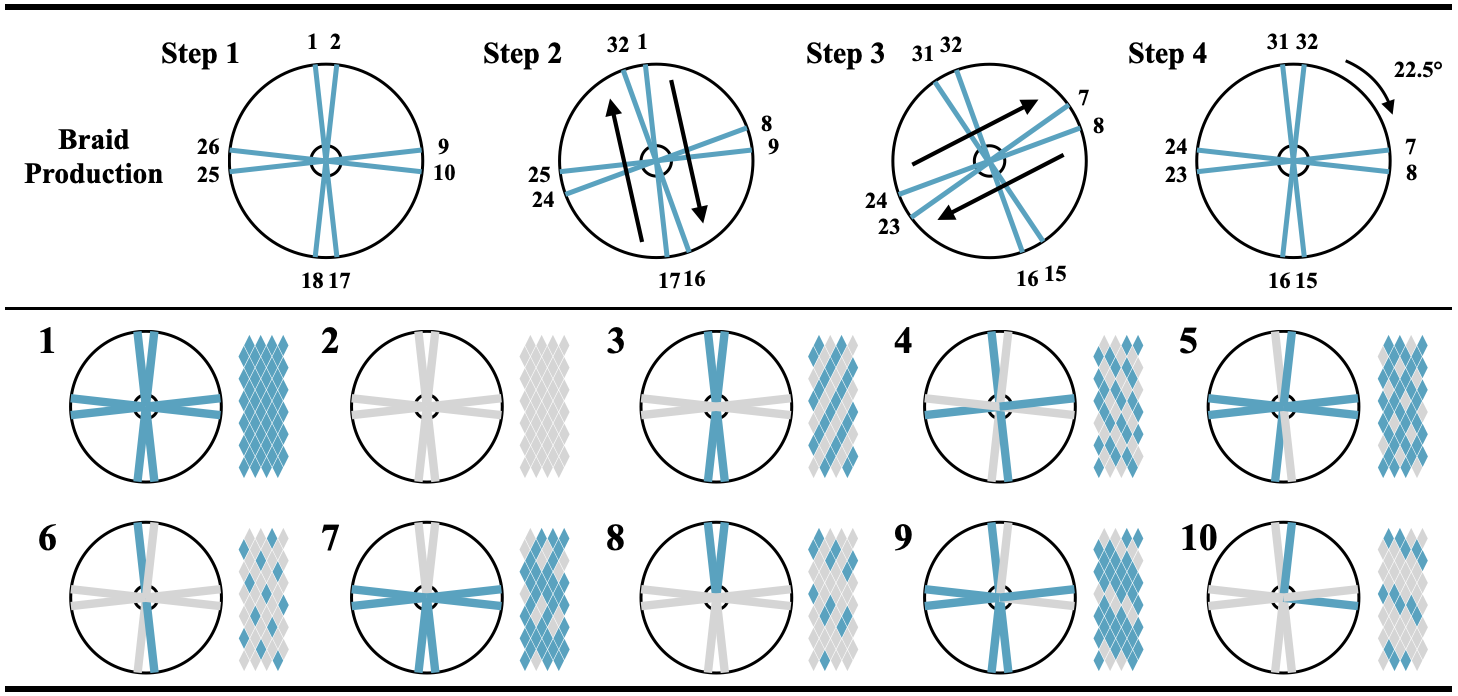}
\caption{The method of preparing the "Eight Vestiges" traditional craft braid using a braid disk and changing the initial position of yarns to create different patterns}
\caption*{Note. Pattern Coding: Pattern 1 is solid cyan, and Pattern 2 is solid white. Patterns 3 and 4 have a cyan-to-white ratio of 1:1; Patterns 5, 7, and 9 have a cyan-to-white ratio of 3:1; and Patterns 6, 8, and 10 have a cyan-to-white ratio of 1:3.}
\label{fig:1}
\end{figure}

    The proper utilization of the available natural resources and waste is crucial for developing sustainability in industry \cite{al-oqlaNaturalFiberReinforced2014}. Therefore, this study focused on ramie which is a plant widely distributed in China and is also known as "Chinese Grass". Its bast fiber, known as ramie fiber, is characterized by high strength, toughness, biodegradability, and cost-effectiveness. Ramie fibers are widely used in fabric production and bio-composites \cite{zhuRecentAdvancesBast2024}. In this study, we used ramie fiber yarns with a diameter of approximately 1 mm to make the braid. To represent the patterns of the braid, cyan and white ramie fiber yarns were selected. Cyan was chosen due to its association with cold colors, evoking feelings of calmness and tranquility. Additionally, white has a clean image and has the effect of reducing one's own presence and emphasizing others \cite{kitagawaPOLYSEMYMATEXT2013}. As a result, participants were expected to perform impression evaluations in a calm and objective manner. There are 10 different patterns that were produced by changing the initial position of the white and cyan yarns as shown in Figure~\ref{fig:1}.

    Furthermore, polymeric materials have replaced conventional materials in industry due to their light weight, ease of processing, and low cost \cite{liuLightweightDesignComposite2016}. To keep the scope of the study manageable, three commonly used polymeric matrix materials were selected: epoxy resin, one-component modified rubber and two-component liquid silicone rubber. These materials have different properties and are widely used in industry. For instance, they have different degrees of transparency, strength, and softness. More importantly, due to their applicability, they are extensively used by designers for model making. Designers can produce composites by adding braids into the matrix material before it cures. Subsequently, the braids of 10 patterns were combined with the matrix materials, producing 30 braid composites.

\begin{figure}[h]
\centering
\includegraphics[width=0.9\textwidth]{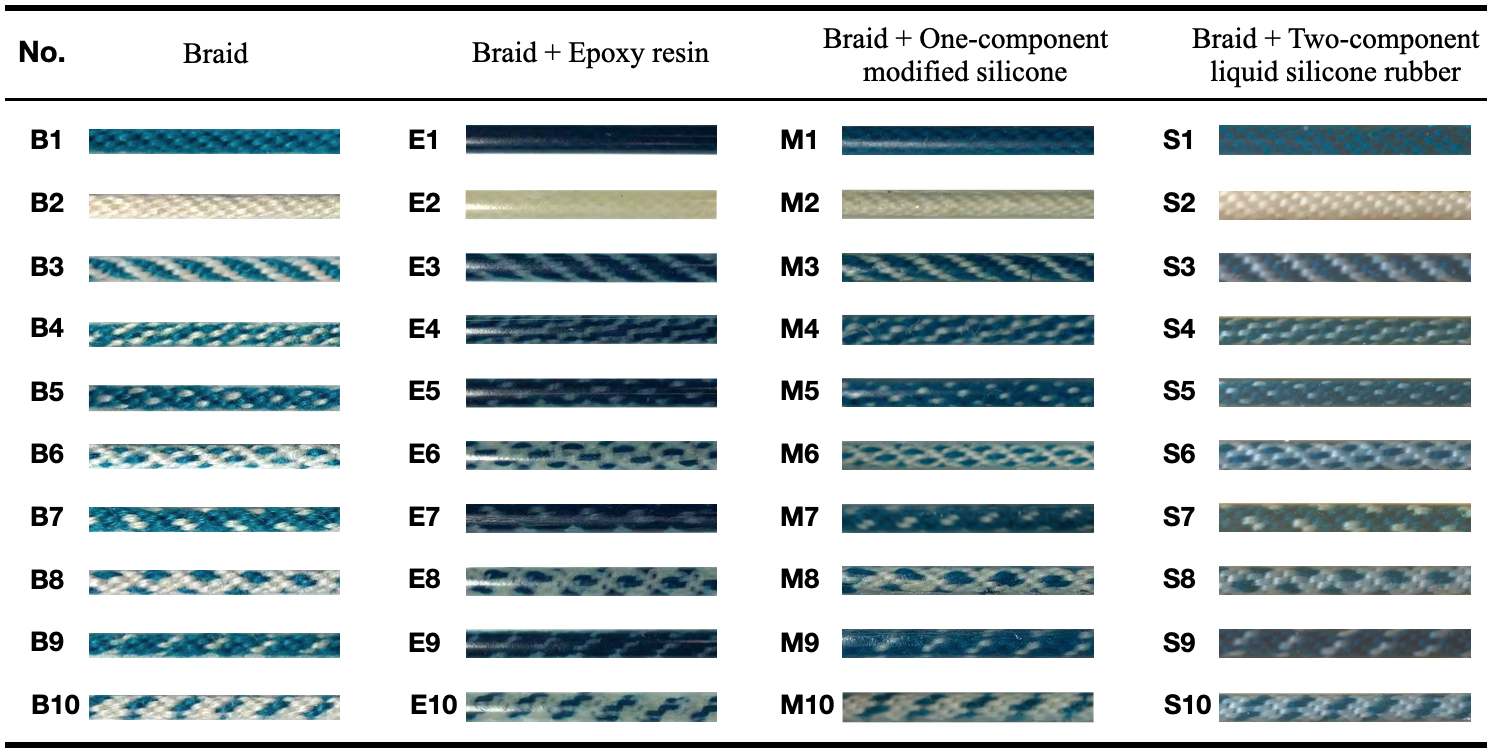}
\caption{Impression evaluation experiment stimuli: A summary based on the differences in matrix material types and braid patterns}
\caption*{Note. Stimuli Coding: Braid, including B1-B10; Braid and epoxy composites, including E1-E10; Braid and one-component modified rubber composites, including M1-M10; Braid and two-component liquid silicone rubber composites, including S1-S10. In this context, B, E, M, and S represent the first letters of Braid, Epoxy, Modified (for modified rubber), and Silicone (for silicone rubber), respectively. The number following each letter corresponds to the braid pattern code, in alignment with the sequence shown in Figure~\ref{fig:1}.}
\label{fig:2}
\end{figure}

    A total of 40 materials, including braids and braid composites, were prepared (see Figure~\ref{fig:2}). The braids have a diameter of 2.5 mm, while the braid composites have a diameter of 4 mm.

\subsection{Method}
    Perceptual learning provides a crucial basis for human expertise that profoundly affect tasks from the pickup of minute sensory detail to the extraction of complex and abstract relations \cite{kellmanPerceptualLearningHuman2009}. To structure the understanding of material experience related to visual and tactile sensations, we adopt the "experience levels" framework for material experiences \cite{karanaMaterialDrivenDesign2015}, and assign specific meanings to these levels based on our research. This represents our attempt to explore materials and construct a model through sensorial, interpretive, affective, and performative levels. In our study, the sensorial level represents the basic impression evaluation items, which can be directly perceived through observation or touch and carry a singular meaning. The interpretive level consists of impression evaluation factors that cannot be directly perceived and encapsulate the composite meanings derived from the basic impression evaluation items. The affective level is a complex factor. We focus on the affective level of "Attractiveness of the Material" and express it through performative level. The performative level includes three ultimate impression evaluation items: "Dislike-Like", "Boring-Interesting" and "Cheap-Luxury". These levels affect each other in a non-sequential manner. Since the sensorial and performative levels can be collected through questionnaire, they are treated as observed variables; whereas the interpretive and affective levels, due to their complexity, are not directly observable and are interpreted through the sensorial and performative levels, thus serving as latent variables.
   
    This study addresses the following research questions: (1) whether matrix materials and patterns significantly affect the performative layer, and (2) whether latent interpretive levels significantly affect the latent affective variable. To address these questions, we conducted a correlation analysis to explore the data and performed a two-way ANOVA to confirm the significant effects of matrix materials and patterns on performative level. Furthermore, based on the results of exploratory factor analysis, a structural equation model was constructed to examine the relationship between latent variables. Data were collected using the 7-point semantic differential method, and the analysis was conducted using IBM SPSS Statistics V26.0 and IBM SPSS AMOS V.27.

\subsection{Impression Evaluation Item}
    Having a detailed and reliable vocabulary is important for meaningfully describing perceptual experience \cite{guestDevelopmentValidationSensory2011}. Accordingly, the impression evaluation items were selected with reference to research on material patterns, textures, as well as visual and tactile perception \cite{guestAudiotactileInteractionsRoughness2002, sakamotoExploringTactilePerceptual2017a, zuoSensoryPerceptionMaterial2016}. Additionally, due to the uniqueness of the stimulus, participants were requested to freely describe the stimulus using adjectives based on their visual and tactile perceptions. A total of 245 freely described adjectives were collected. These synonymous adjectives were then combined, and appeared more than five times were selected as impression evaluation items. Based on the references and freely descriptions, 22 basic impression evaluation items were selected for the impression evaluation experiments. Furthermore, "Dislike-Like", "Boring-Interesting" and "Cheap-Luxury" were used as ultimate impression evaluation items, serving as explanatory variables for the "Attractiveness of the material". As shown in Table~\ref{tab:1}, a total of 25 items were used in the impression evaluation experiment. In this experiment, we used the 7-point semantic differential method to collect these impression evaluation items.

\begin{table}[htbp]
\centering
\caption{Impression evaluation items selected based on references and freely describe}
\begin{tabular}{lll}
\toprule
Basic Impression (Sensorial Level) & & \\
\midrule
Hard - Soft & Wet - Dry & Old - New \\
Dirty - Clean & Light - Heavy & Cold - Warm \\
Dark - Bright & Weak - Strong & Cloudy - Clear \\
Elasticity - Plasticity & Fragile - Sturdy & Traditional - Modern \\
Rustic - Fancy & Disorderly - Orderly & Rough - Smooth \\
Bad Touch - Good Touch & Simple - Complex & Plant Feeling - Animal Feeling \\
Vulgar - Elegant & Low Elasticity - High Elasticity & Gloomy - Shiny \\
Unapproachable - Approachable & & \\
\midrule
Ultimate Impression (Performative Level) & & \\
\midrule
Dislike - Like & Boring - Interesting & Cheap - Luxury \\
\bottomrule
\end{tabular}
\label{tab:1}
\end{table}

\subsection{Data Processing and Analysis}
    Step 1. Correlation Analysis: Pearson correlation analysis was conducted to explore the data structure. When multiple pairs of impression evaluation items showed strong correlations, we assessed whether dimensionality reduction methods could be applied to analyze the data.
    
    Step 2. Two-way ANOVA: Two-way ANOVA was conducted to analyze whether differences in matrix materials and patterns had a significant effect on performative levels, as well as to examine the presence of interaction effects. The main effects and interaction effect sizes were estimated.

    Step 3. Exploratory Factor Analysis (EFA): Exploratory factor analysis was conducted using the maximum likelihood method with Promax rotation to determine the optimal number of factors and identify the impression evaluation items included in each factor. Eigenvalues greater than 1.00 were used as the criterion for retaining factors, and impression evaluation items with factor loadings greater than 0.60 were retained.

    Step 4. Structural Equations Modeling (SEM): Structural Equation Modeling combines factor analysis and path analysis to assess complex models involving multiple linear equations. It also supports model comparisons by evaluating the fit of alternative models, helping to identify the best model \cite{hairPartialLeastSquares2021}. Based on the results of the exploratory factor analysis, the maximum likelihood method was used to construct the structural equation model. We examined the correlations between the interpretive levels and compared the regression coefficients of the interpretive level on the affective level. Seven commonly used fit indices were selected to evaluate model fit \cite{iacobucciStructuralEquationsModeling2010, jacksonReportingPracticesConfirmatory2009, sahooStructuralEquationModeling2019a}: the normalized chi-square index ($\chi^2$/df), the goodness of fit index (GFI), the adjusted goodness of fit index (AGFI), the comparative fit index (CFI), the Tucker-Lewis index (TLI), the root mean square error of approximation (RMSEA), and the standardized root mean square residual (SRMR). An satisfactory model fit is defined by the following criteria: $\chi^2$/df $\leq$ 3.00, GFI, AGFI and CFI $\geq$ 0.90, TLI $\geq$ 0.95, RMSEA and SRMR $\leq$ 0.08.

\subsection{Participants and experimental environment}
    In this experiment, 10 university students (aged 20-30 years, with an average age of 25 years) who had completed at least three years of design education participated in the impression evaluation experiment. The participants had professional material experience and a comprehensive understanding of the impression evaluation items, beyond just a literal understanding. They evaluated the impressions of the stimulus without being informed of the purpose of the experiment. Participants were allowed to observe and touch the stimulus at their own pace, without time constraints. Prior to the experiment, all stimulus were presented to the participants for familiarization and to assess the relationships between the stimulus. Additionally, to minimize the potential influence of the experimental environment on the results, the experiment was conducted under controlled conditions: room temperature of $25^\circ\mathrm{C}\pm1^\circ\mathrm{C}$, humidity of $30\%\pm10\%$, illumination of $300$--$500~\mathrm{lx}$, color temperature of $3500$--$4000~\mathrm{K}$.

\section{Results}
\subsection{Correlation Analysis}
\begin{figure}[h]
    \centering
    \includegraphics[width=0.9\textwidth]{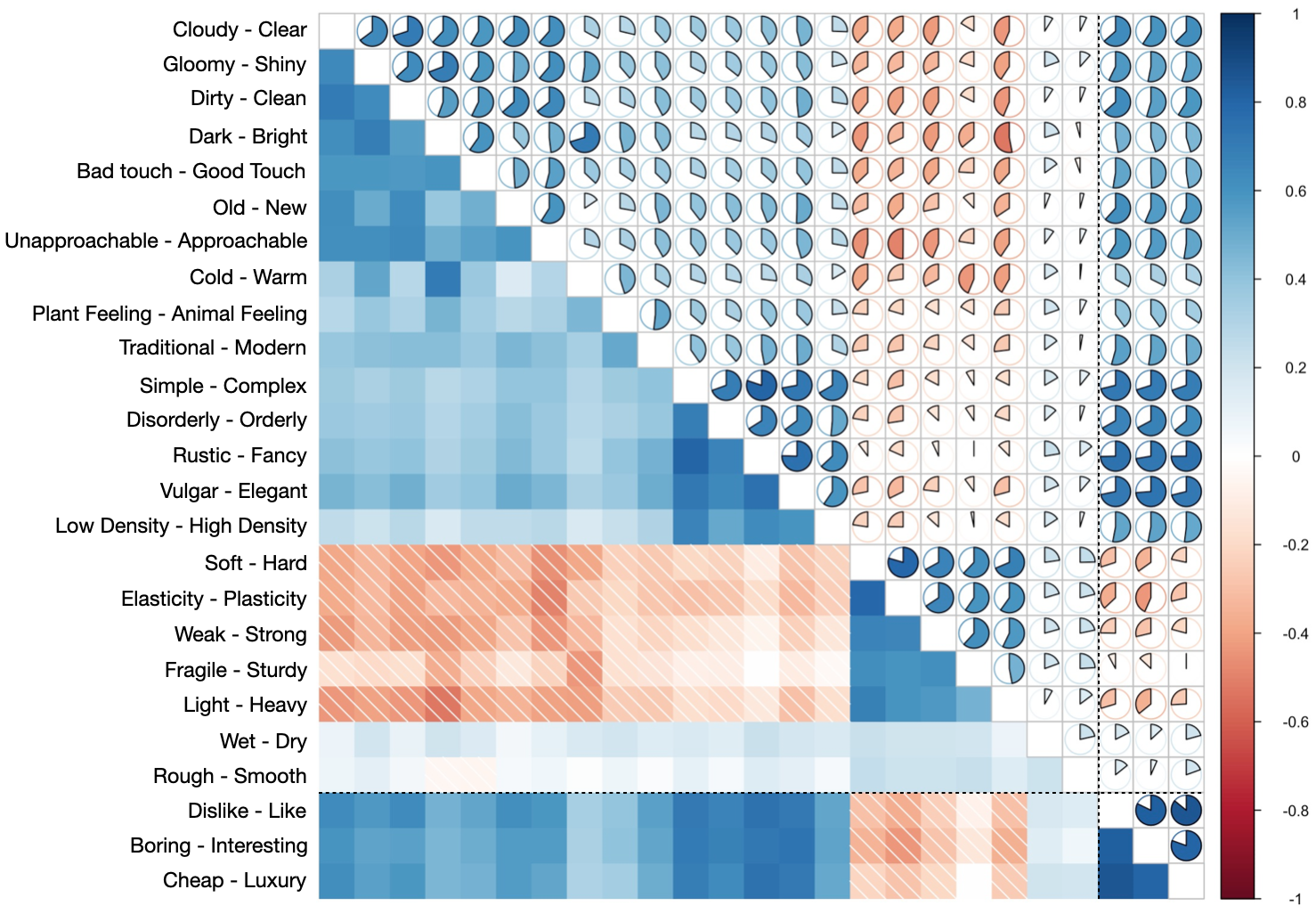}
    \caption{Correlation heatmap of impression evaluation items}
    \caption*{Note. Blue and red represent positive and negative correlations, respectively. Darker shades represent stronger correlations.}
    \label{fig:3}
\end{figure}

    The results of the correlation analysis are shown in Figure~\ref{fig:3}. The findings reveal correlations among several basic evaluation items. Consequently, reducing dimensionality by summarizing highly correlated items into a single factor can improve the interpretability of the model, making it easier to understand. It is worth paying attention to the correlation between the ultimate evaluation items. The correlation coefficients for "Dislike-Like" and "Boring-Interesting" are 0.822 ($p < 0.001$), for "Dislike-Like" and "Cheap-Luxury" are 0.858 ($p < 0.001$), for "Boring-Interesting" and "Cheap-Luxury" are 0.805 ($p < 0.001$), all these correlation coefficients are statistically significant which indicates that there is a high degree of correlation between the performative level. Therefore, the observation variables can provide a strong explanation for the "Attractiveness of the Material".

\subsection{Two-ways ANOVA}
\begin{figure}[h]
    \centering
    \includegraphics[width=0.9\textwidth]{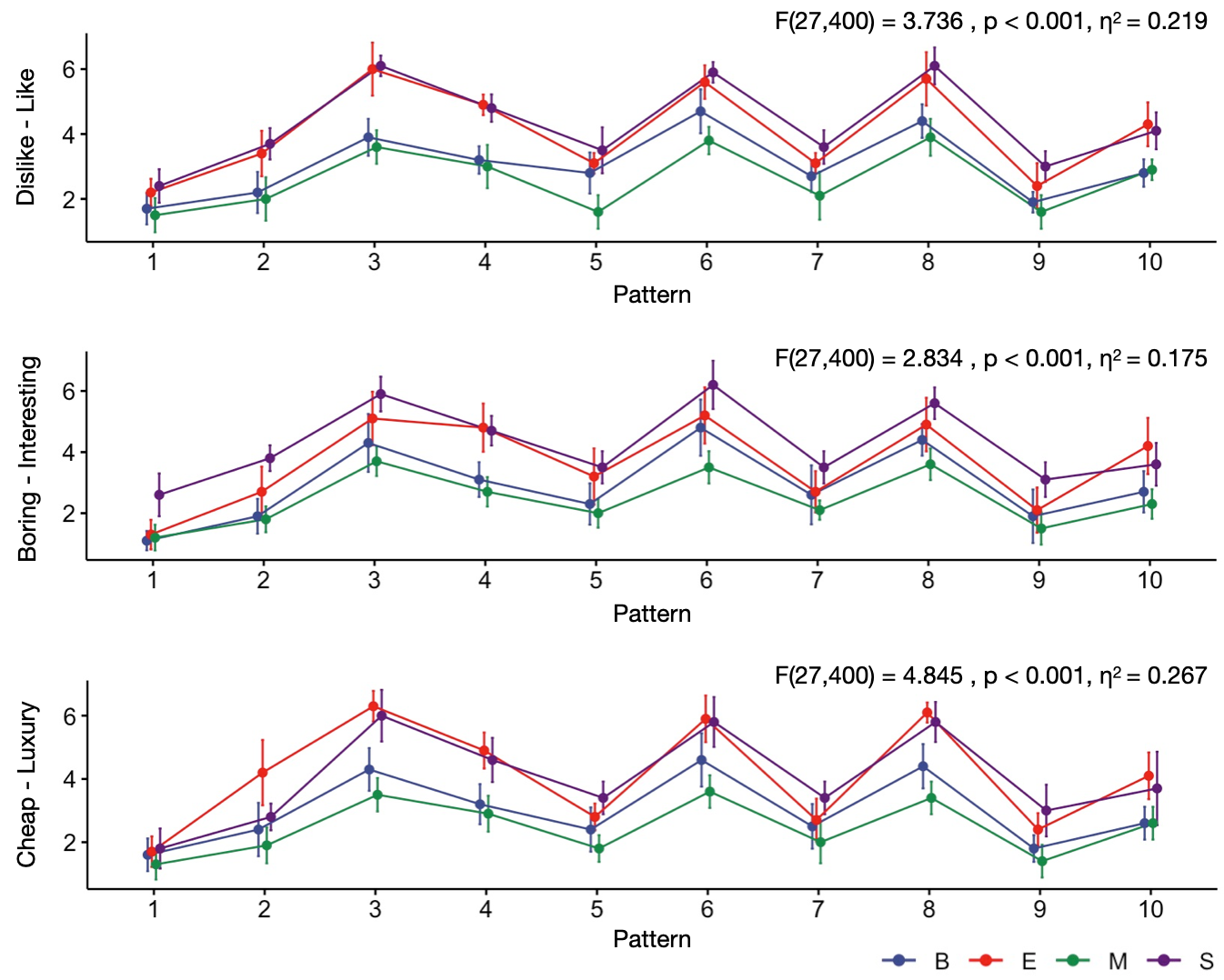}
    \caption{Main and interaction effect of matrix materials and patterns on "Dislike-Like", "Boring-Interesting", and "Cheap-Luxury" were examined, with means and standard deviations provided}
    \label{fig:4}
\end{figure}

    The results of the two-way ANOVA are shown in Figure~\ref{fig:4}, indicating that the difference in matrix materials have a significant effect on "Dislike-Like" ($F(3, 400) = 227.026$, $p < 0.001$, $\eta^2 = 0.654$), "Boring-Interesting" ($F(3, 400) = 143.312$, $p < 0.001$, $\eta^2 = 0.544$), and "Cheap-Luxury" ($F(3, 400) = 160.961$, $p < 0.001$, $\eta^2 = 0.573$). Similarly, difference in patterns also significantly effect on "Dislike-Like" ($F(9, 400) = 182.067$, $p < 0.001$, $\eta^2 = 0.820$), "Boring-Interesting" ($F(9, 400) = 124.582$, $p < 0.001$, $\eta^2 = 0.757$), and "Cheap-Luxury" ($F(9, 400) = 150.915$, $p < 0.001$, $\eta^2 = 0.790$). Furthermore, the interaction effect between matrix materials and patterns significantly effects on "Dislike-Like" ($F(27, 400) = 3.736$, $p < 0.001$, $\eta^2 = 0.219$), "Boring-Interesting" ($F(27, 400) = 2.834$, $p < 0.001$, $\eta^2 = 0.175$), and "Cheap-Luxury" ($F(27, 400) = 4.845$, $p < 0.001$, $\eta^2 = 0.267$). The average values of the performative level show that combining matrix materials such as epoxy resin and two-component liquid silicone rubber leads to an increase in "Dislike-Like", "Boring-Interesting" and "Cheap-Luxury", whereas combining with one-component modified rubber results in a decrease.

\subsection{Exploratory Factor Analysis (EFA)}
    Exploratory factor analysis was conducted to determine the number of factors and examine the correlations between each factor and the observed variables. The results showed that the KMO value was 0.92, and Bartlett's test of sphericity statistic was 4722.27 with a p-value close to 0.00, indicating that the data were appropriate for factor analysis. As shown in Table~\ref{tab:2}, three factors were extracted with eigenvalues greater than 1.00, specifically 7.12, 2.49, and 1.32. These three factors explained a total variance of 64.26

\begin{table}[htbp]
\centering
\caption{Factor analysis results: Factor loadings and variance explained}
\begin{tabular}{l r r r r r r}
\toprule
Impression Evaluation Item & Factor 1 & Factor 2 & Factor 3 & Extraction & Mean & SD \\
\midrule
Gloomy - Shiny & .873 & -.048 & .083 & .651 & 3.57 & 1.39 \\
Cloudy - Clear & .862 & -.006 & .039 & .702 & 3.98 & 1.39 \\
Dirty - Clean & .810 & .018 & .002 & .671 & 3.98 & 1.33 \\
Dark - Bright & .746 & -.099 & -.090 & .563 & 3.67 & 1.85 \\
Bad Touch - Good Touch & .702 & .005 & -.030 & .520 & 3.99 & 1.23 \\
Old - New & .628 & .157 & .025 & .507 & 3.54 & 1.21 \\
Unapproachable - Approachable & .624 & .084 & -.147 & .578 & 3.69 & 1.03 \\
\midrule
Simple - Complex & -.078 & .933 & -.032 & .811 & 3.45 & 1.45 \\
Rustic - Fancy & .122 & .873 & .180 & .824 & 3.49 & 1.62 \\
Low Density - High Density & -.170 & .789 & -.104 & .539 & 3.73 & 1.25 \\
Vulgar - Elegant & .156 & .741 & -.024 & .711 & 3.52 & 1.46 \\
Disorderly - Orderly & .070 & .716 & -.011 & .577 & 3.78 & 1.34 \\
\midrule
Hard - Soft & .017 & -.033 & .910 & .827 & 4.42 & 1.44 \\
Elasticity - Plasticity & .061 & -.149 & .858 & .758 & 4.51 & 1.39 \\
Fragile - Sturdy & .124 & .046 & .781 & .513 & 4.20 & 1.10 \\
Weak - Strong & -.163 & .079 & .701 & .606 & 4.41 & 1.05 \\
Light - Heavy & -.255 & .064 & .607 & .566 & 4.26 & .93 \\
\midrule
Eigenvalue & 7.12 & 2.49 & 1.32 &       &       &       \\
Cumulative (\%) & 41.87 & 14.65 & 7.74 &       &       &       \\
Cumulative Contribution Ratio (\%) & 41.87 & 56.52 & 64.26 &       &       &       \\
\bottomrule
\end{tabular}
\caption*{Note. Exploratory factor analysis was conducted using the maximum likelihood method with Promax rotation.}
\label{tab:2}
\end{table}

    The items in the first factor (7 items) are primarily related to the intrinsic properties of the matrix material, so this factor was named "Intrinsic Property". The items in the second factor (5 items) are mainly associated with the pattern of the reinforcement material, and this factor was named "Aesthetic Property". The items in the third factor (5 items) are mainly concerned with the physical properties of the composite, the third factor was named "Physical Property". Since the factor loadings were below 0.6, the following items were excluded: "Cold-Warm" (M = 3.47, SD = 1.39), "Plant Feeling-Animal Feeling" (M = 3.98, SD = 0.97), "Traditional-Modern" (M = 3.82, SD = 0.90), "Wet-Dry" (M = 3.96, SD = 0.83), and "Rough-Smooth" (M = 4.20, SD = 0.81). In addition, the extraction, mean, and standard deviation of the data were calculated (see Table~\ref{tab:2}).

\subsection{Structural Equations Modeling (SEM)}
\begin{figure}[h]
    \centering
    \includegraphics[width=\textwidth]{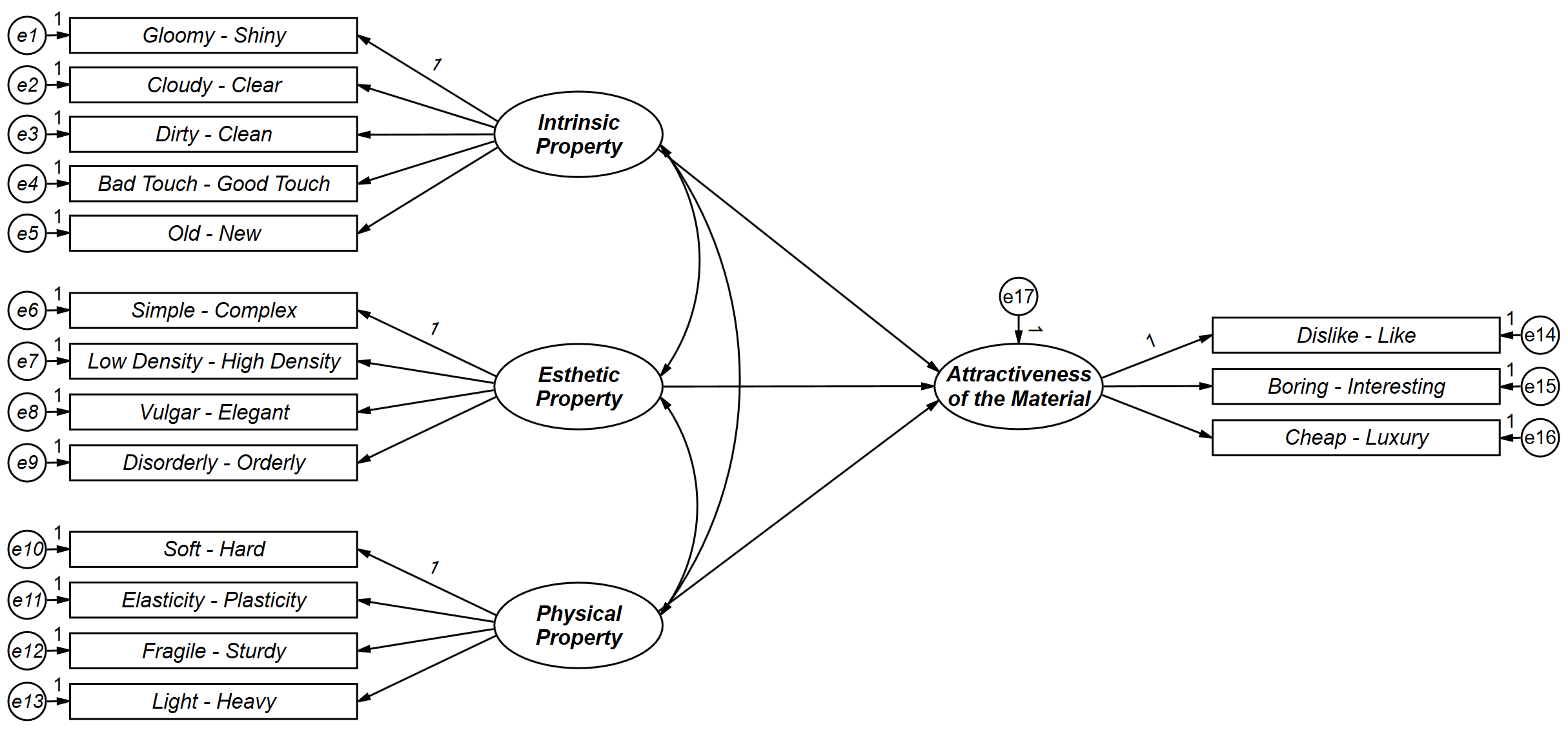}
    \caption{Structural equation modeling of braids and braid composites based on visual-tactile sensory properties: Path diagram}
    \label{fig:5}
\end{figure}

    The structural equation model was constructed based on the results of the exploratory factor analysis, and was subsequently modified to improve the fit indices. The model was finalized after removing the items "Dark-Bright", "Unapproachable-Approachable", "Rustic-Fancy" and "Weak-Strong", as shown in Figure~\ref{fig:5}. Table~\ref{tab:3} show the fit indices of the structural equation model, indicating that the model demonstrates a good fit.

\begin{table}[htbp]
  \centering
  \caption{Fit indices, cutoff values, analytical results, and model evaluation in structural equation modeling}
  \resizebox{\textwidth}{!}{%
  \begin{tabular}{lccccccc}
    \toprule
    & $\chi^2$/df & GFI & AGFI & CFI & TLI & RMSEA & SRMR \\
    \midrule
    Cutoff value & $<$ 3.00 & $>$ 0.90 & $>$ 0.90 & $>$ 0.90 & $>$ 0.95 & $<$ 0.08 & $<$ 0.08 \\
    Analysis Results & 2.684 & .925 & .896 & .965 & .958 & .065 & .057 \\
    Model Evaluation & Good fitting & Good fitting & Accept & Good fitting & Good fitting & Good fitting & Good fitting \\
    \bottomrule
  \end{tabular}%
  }
  \label{tab:3}
\end{table}

    The results of the structural equation model are shown in Table~\ref{tab:4}. There are five impression evaluation items in the first factor "Intrinsic Property" with factor loading of "Gloomy-Shiny" is 0.759, "Cloudy-Clear" is 0.847, "Dirty-Clean" is 0.829, "Bad Touch-Good Touch" is 0.700, "Old-New" is 0.736, with composite reliability (CR) of 0.883 and average variance extracted (AVE) of 0.602. There are four impression evaluation items in the second factor "Aesthetic Property" with factor loading of "Simple-Complex" is 0.879, "Low Density-High Density" is 0.701, "Vulgar-Elegant" is 0.839, "Disorderly-Orderly" is 0.784, with composite reliability of 0.879 and average variance extracted of 0.646. There are four impression evaluation items in the third factor "Physical Property" with factor loading of "Hard-Soft" is 0.928, "Elasticity-Plasticity" is 0.856, "Fragile-Sturdy" is 0.667, "Light-Heavy" is 0.743, with composite reliability of 0.877 and average variance extracted of 0.645. From the affective level perspective, in the factor "Attractiveness of the Material" with factor loading of "Dislike-Like" is 0.935, "Boring-Interesting" is 0.886, "Cheap-Luxury" is 0.911, with composite reliability of 0.936 and average variance extracted of 0.830. According to Fornell and Larcker's \cite{fornellEvaluatingStructuralEquation1981} recommendations, a composite reliability (CR) greater than 0.7 is considered acceptable, and the average variance extracted (AVE) should be greater than 0.5 to ensure acceptable convergent validity. Since each factor in the constructed structural equation model satisfies this criterion, we conclude that the model demonstrates excellent convergent validity.

\begin{table}[htbp]
\centering
\caption{Summary of structural equation model parameters and convergent validity}
\begin{tabular}{l l r r r r r r r r}
\toprule
Factors & Item & Un-std. & S.E. & t-value & P & Std. & SMC & CR & AVE \\
\midrule
Intrinsic Property & Gloomy - Shiny         & 1.000 &       &        &      & .759 & .576 &       &       \\
                   & Cloudy - Clear         & 1.120 & .064  & 17.583 & ***  & .847 & .717 &       &       \\
                   & Dirty - Clean          & 1.044 & .061  & 17.166 & ***  & .829 & .687 & .883  & .602  \\
                   & Bad Touch - Good Touch & .817  & .058  & 14.175 & ***  & .700 & .490 &       &       \\
                   & Old - New              & .847  & .057  & 14.984 & ***  & .736 & .542 &       &       \\
\midrule
Aesthetic Property & Simple - Complex       & 1.000 &       &        &      & .879 & .773 &       &       \\
                   & Low - High Density     & .684  & .042  & 16.379 & ***  & .701 & .491 & .879  & .646  \\
                   & Vulgar - Elegant       & .958  & .044  & 21.955 & ***  & .839 & .704 &       &       \\
                   & Disorderly - Orderly   & .824  & .042  & 19.537 & ***  & .784 & .615 &       &       \\
\midrule
Physical Property  & Soft - Hard            & 1.000 &       &        &      & .928 & .861 &       &       \\
                   & Elasticity - Plasticity& .892  & .039  & 23.164 & ***  & .856 & .733 & .877  & .645  \\
                   & Fragile - Sturdy       & .548  & .035  & 15.525 & ***  & .667 & .445 &       &       \\
                   & Light - Heavy          & .511  & .028  & 17.970 & ***  & .734 & .539 &       &       \\
\midrule
Attractiveness     & Dislike - Like         & 1.000 &       &        &      & .935 & .874 &       &       \\
of the Material    & Boring - Interesting   & .976  & .033  & 29.648 & ***  & .886 & .785 & .936  & .830  \\
                   & Cheap - Luxury         & 1.053 & .033  & 32.275 & ***  & .911 & .830 &       &       \\
\bottomrule
\end{tabular}
\caption*{Note. Un-std: Unstandardized parameter estimates; S.E.: Standard Error; Std.: Standardized parameter estimates; SMC: Squared Multiple Correlation; CR: Composite Reliability; AVE: Average Variance Extracted}
\label{tab:4}
\end{table}

    In addition, the Fornell-Larcker criterion was used to assess discriminant validity, based on the logic that a factor should share more variance with its associated indicators than with any other factor \cite{fornellEvaluatingStructuralEquation1981}. The discriminant validity between factors is shown in Table~\ref{tab:5}. The results indicate that the factors across the interpretive levels demonstrate strong discriminant validity. Moreover, we also found that "Intrinsic Property" were positively correlated with "Aesthetic Property" with a correlation coefficient of 0.574 ($p < 0.001$), while "Intrinsic Property" were negatively correlated with "Physical Property" with a correlation coefficient of -0.520 ($p < 0.001$), and "Aesthetic Property" were negatively correlated with "Physical Property" with a correlation coefficient of -0.326 ($p < 0.001$).

\begin{table}[htbp]
\centering
\caption{Factor correlation coefficients and discriminant validity}
\begin{tabular}{l r c c c c}
\toprule
Factor & AVE &
  \begin{tabular}{@{}c@{}}Intrinsic\\Property\end{tabular} &
  \begin{tabular}{@{}c@{}}Aesthetic\\Property\end{tabular} &
  \begin{tabular}{@{}c@{}}Physical\\Property\end{tabular} &
  \begin{tabular}{@{}c@{}}Attractiveness of\\the Material\end{tabular} \\
\midrule
Intrinsic Property              & \phantom{0}.602 &  \phantom{0}.776 &       &       &       \\
Aesthetic Property              & \phantom{0}.646 &  \phantom{0}.574 &  \phantom{0}.804 &       &       \\
Physical Property               & \phantom{0}.645 & -.520           & -.326           &  \phantom{0}.803 &       \\
Attractiveness of the Material  & \phantom{0}.830 &  \phantom{0}.806 &  \phantom{0}.895 & -.362           &  \phantom{0}.911 \\
\bottomrule
\end{tabular}
\label{tab:5}
\end{table}

    The results of the path analysis are shown in Table~\ref{tab:6}. The standardized regression coefficients from "Intrinsic Property", "Aesthetic Property", and "Physical Property" to the "Attractiveness of the Material" are 0.486 ($p < 0.001$), 0.650 ($p < 0.001$), and 0.103 ($p < 0.001$), respectively. These results indicate that all three factors have a significant positive effect on the "Attractiveness of the Material".

\begin{table}[htbp]
\centering
\caption{Summary of path analysis parameters and estimates}
\begin{tabular}{llrrrrr}
\toprule
Path Analyses & & Un-std. & S.E. & t-value & P & Std. \\
\midrule
Intrinsic Property & & .621 & .053 & 11.715 & *** & .486 \\
Aesthetic Property & $\to$ Attractiveness of the Material & .685 & .038 & 17.793 & *** & .650 \\
Physical Property  & & .104 & .029 & 3.523  & *** & .103 \\
\bottomrule
\end{tabular}
\caption*{Note. Un-std: Unstandardized parameter estimates; S.E.: Standard Error; Std.: Standardized parameter estimates}
\label{tab:6}
\end{table}

\section{Discussion}
    Additional interviews with participants were conducted to further discuss the findings in the context of the models. The research findings indicate that the intrinsic properties are mainly related to the visual and tactile perception of the matrix material. The shinier, clearer, and more tactilely pleasing the matrix material, the higher its intrinsic property score. However, braids, which lack a matrix material, present a challenge for identifying intrinsic properties. Despite this, the strong correlation between intrinsic and aesthetic properties suggests that the intrinsic properties of braids are influenced by their aesthetic properties to some extent. The patterns are more strongly associated with visual stimulation, the higher the complexity and regularity of the pattern, the higher the aesthetic properties score. This finding is consistent with Lazard and Mackert's \cite{lazardUserEvaluationsDesign2014} research, which states that increased design complexity is associated with higher levels of perceived design aesthetics. In addition, physical properties are primarily identified through tactile perception, with higher strength and lower tensile deformation corresponding to higher physical property scores.
    
    However, in design practice, higher scores do not necessarily correspond to better performance. For example, materials with gloomy or cloudy characteristics may offer better privacy, despite scoring lower on intrinsic properties. Similarly, materials with low strength and high deformation may score lower on physical properties, but are commonly used in applications such as impact absorption. The correlation analysis also revealed a strong positive correlation between intrinsic and aesthetic properties, as well as a strong negative correlation between intrinsic and physical properties. Therefore, designers can control the type of matrix material and the pattern of the braid to modify the material properties for specific design purposes. For instance, materials with high intrinsic and aesthetic properties but low physical properties tend to be more conspicuous, while the opposite is true for materials that are less noticeable. This was further confirmed in the interviews, where most participants emphasized the importance of quantitative research on the perceived properties of materials, while also noting that the specific use context, including its purpose and environment, must be taken into account.
    
    The significant difference in the sensory experience of braids and braid composites has been demonstrated through composite process. Specifically, the use of braids compounded with epoxy resin, one-component modified rubber and two-component liquid silicone rubber leads to significant differences at the performative level, which in turn influences the attractiveness of the material. Therefore, designers should pay attention to using the changes in material properties induced by the composite process to optimize the design. Furthermore, in conjunction with the earlier study on physical properties \cite{zhangEffectDifferentMatrix2024}, it was found that the physical properties of composite change considerably depending on the matrix material. Compared to braids, composites made of braid and epoxy exhibit high strength and low deformation, whereas composites made of braid and one-component modified rubber, as well as braid and two-component liquid silicone rubber, are characterized by high deformation and low strength. Additionally, the physical properties of these composites are more stabilized. This is because, when combined with the matrix material, it helps transfer stress, resulting in a more uniform distribution of forces across the fibers. Although physical properties have a limited effect on sensory evaluation, they directly influence the practical applications of material. Therefore, designers need to rationally select the physical properties of materials based on design requirements, and consider the use of composites to enhance physical performance when a single material is insufficient to meet the design needs. Moreover, in the context of integrating design and engineering, the use of braids has also contributed to the spread of traditional culture. In summary, the process of combining braids with matrix materials to develop creative materials plays a crucial role in design, engineering, and the dissemination of traditional craftsmanship.
    
    In the field of design, CMF (Color, Material, Finish) remains a prominent and enduring research direction. While the importance of aesthetic attributes is widely recognized, functionality remains a priority from a broader perspective. Therefore, the functionality of materials should be prioritized based on the intended use before considering aesthetic aspects, following the design principle of "form follows function'', to achieve a harmonious balance between technological performance and user experience.

\section{Limitations and research perspectives}
    Due to the limited availability of participants with specialized material experience, a small sample size study was conducted. In addition, the rule of thumb that the sample size should exceed 100, and preferably 10 to 20 times the number of observed variables, was taken into consideration \cite{andersonStructuralEquationModeling1988, christopherwestlandLowerBoundsSample2010, tinsleyUsesFactorAnalysis1987}. To enhance the generalizability of the model, future studies will involve a larger and more diverse participant pool. Moreover, multiple group comparisons will be conducted based on factors such as gender and nationality.
    
    The selection of stimuli in this study represents only a subset of braids and braid composites, although efforts were made to capture the diversity of material properties and their representativeness. In addition, we have considered that the three matrix materials are easy to handle and can be readily applied according to the designer's needs. Therefore, the findings can serve as a useful reference for designers in selecting creative materials. However, future research will aim to include a broader range of representative materials.

\section{Conclusions}
    In this study, ramie fiber braids were compounded with epoxy resin, one-component modified rubber, and two-component liquid silicone rubber to prepare creative materials. The effect of these materials on impression evaluation was analyzed and compared. It was found that compounding with the matrix material significantly altered the sensory properties of creative materials. Specifically, compared to the braid, the braid and one-component modified rubber composite showed a decrease in attractiveness, as reflected by lower scores on the "Dislike-Like", "Boring-Interesting", and "Cheap-Luxury". In contrast, the braid and epoxy resin composite, as well as the braid and two-component liquid silicone rubber composite, achieved the highest scores on the "Dislike-Like", "Boring-Interesting", and "Cheap-Luxury", respectively, leading to increased attractiveness. More importantly, the study proposes a structural equation model for evaluating the attractiveness of braids and braid composites. Structural equation modeling indicates that the "Intrinsic Property", "Aesthetic Property", and "Physical Property" significantly affect "Attractiveness of the Material". Regression coefficient comparisons revealed that aesthetic properties had the strongest effect, followed by intrinsic properties, while physical properties had a weaker effect on attractiveness. Therefore, it can be concluded that a matrix material with higher gloss, clearer appearance, and better tactile sensation is more attractive when compounded with a braid featuring a complex and ordered pattern. Moreover, in the field of design, matrix material should be selected based on the intended use, and the pattern should be selected according to the environmental context to maximize the characteristics of the material.

\section{Statements and Declarations}
    The authors have no competing interests to declare that are relevant to the content of this article.

\bibliographystyle{unsrt}
\bibliography{references}






\end{document}